# Third-Party Credit Guarantees and the Cost of Debt: Evidence from Corporate Loans*

**Mehdi Beyhaghi**

## Abstract

Using a comprehensive dataset collected by the Federal Reserve, I find that over one-third of corporate loans issued by US banks are fully guaranteed by legal entities separate from borrowing firms. Using an empirical strategy that accounts for time-varying firm and lender effects, I find that the existence of a third-party credit guarantee is negatively related to loan risk, loan rate, and loan delinquency. Third-party credit guarantees alleviate the effect of collateral constraints in credit market. Firms (particularly smaller firms) that experience a negative shock to their asset values are less likely to use collateral and more likely to use credit guarantees in new borrowings.

## 1. Introduction

A body of theoretical literature focuses on the importance of the asset value channel (or the "collateral channel") of bank lending for business cycles.[1] Through this channel, temporary shocks to asset values affect agents' borrowing capacity and cost of debt, which in turn can have long-lasting effects on the firm's investments and economic output. In line with this literature, a number of recent studies empirically examine the relation between the values of assets owned by a borrowing firm and the characteristics of the loan contract that the firm obtains (Benmelech and Bergman, 2009; Cerqueiro, Ongena, and Roszbach, 2016; Luck and Santos, 2019, among others). These studies find that all else being equal, firms that can

* I thank Amiyatosh Purnanandam (editor), an anonymous co-editor, an anonymous referee, and Azamat Abdymomunov, Christa Bouwman, Jon Bzdawka, Jennifer Conrad, Bob DeYoung, Cesare Fracassi, Jeff Gerlach, Victoria Ivashina, Cooper Killen, Atanas Mihov, and Phil Strahan for discussions and comments on the article. The views expressed in this article are solely those of the author. They do not necessarily reflect the views of the Federal Reserve Bank of Richmond or the Federal Reserve System.

1 See, for example, Bernanke and Gertler (1986), Kiyotaki and Moore (1997), Gorton and Ordõez (2014), and Liu and Sinclair (2020).

pledge assets as collateral find it easier to obtain credit at a reduced cost. On the other hand, firms that experience a negative shock to the value of their assets have difficulty pledging collateral and therefore are not able to obtain credit or obtain credit at an increased cost. The issue is compounded when firms rely on their assets, not only for financing purposes but also to cover future payments to hedging counterparties.[2]

In this study, I test the notion that asset constraint is the key determinant of borrowing capacity for the majority of firms. In particular, I focus on third-party credit guarantee, an arrangement used by many corporate borrowers instead of or in addition to pledging their own assets as collateral. My goal is to test if the widespread use of third-party credit guarantees dilutes the effectiveness of the asset value channel. In the presence of a third-party credit guarantee, a lender is protected from adverse idiosyncratic shocks to a borrower's assets. As a result, the lender is expected to extend a loan more easily and at better terms. My research is motivated by the anecdotal evidence that suggests many borrowing firms, especially smaller and growing borrowers, do not have sufficient high-quality assets to pledge as collateral. Moreover, many borrowing firms are not willing to pledge collateral to have more flexibility to sell or redeploy assets or to use assets for future financing or hedging purposes (Rampini and Viswanathan, 2013; Li, Whited, and Wu, 2016; Benmelech, Kumar, and Rajan, 2020).[3]

To address the research questions in this study, I use a unique and comprehensive panel data-set containing details of all USD commercial and industrial loans with $1 million or more in commitment that are issued by large bank holding companies in the USA.[4] My first finding is that third-party credit guarantees are frequently used in US bank lending: Over 46% of corporate loans (equivalent to over 40% of all bank exposures in USD through corporate loans) are fully or partially guaranteed by a legal entity separate from the borrowing firm.[5] More importantly, over 42% of all corporate loans (equivalent to over 36% of bank exposures in USD through corporate loans) are fully guaranteed in the sense that for each loan, there is "explicit" recourse for full repayment of the credit obligation by a third party.

2 See Rampini and Viswanathan (2010), Rampini and Viswanathan (2013), and Rampini, Sufi, and Viswanathan (2014), who argue that, in addition to affecting borrowing capacity, collateral constraints have implications for risk management.

3 This study also pays attention to three popular forms of third-party credit guarantees (governmental, corporate, and personal) and aims to examine the separate effect of each form on loan contracting.

4 This constitutes over 97% of the dollar value of total USD C&I exposures on the current balance sheets of the thirty largest US banks. The data are collected by the Federal Reserve under Y14Q schedules. I use data from the Corporate Loan Data Schedule (Schedule H.1), which cover detailed information on the status of bank loan facilities and their guarantors for the full set of corporate loans on banks' portfolios since 2012Q2. The data-set also includes bank ratings for borrowing firms as well as bank estimates of each loan's LGD and the performance of the loan over time. The most recent instructions and data fields are publicly available at https://www.federalreserve.gov/reportforms/forms/FR_Y-14Q20181231_i.pdf.

5 Although the source of data does not cover details of small business loans (3% of exposures), the study by Rice and Strahan (2010), on the role of credit competition on small-firm finance, suggests that having a credit guarantee is even more common for small business loans. Using the Survey of Small Business Finance, the authors find that at least 56% of loans to small firms in the USA are guaranteed.

Focusing on new USD loans (about 150,000 loans) during the sample period (2012–18), next, I aim to examine the implications of having a third-party guarantee for corporate borrowing and bank risk. An immediate empirical challenge, however, is that the existence of a credit guarantee is not random and is likely to be determined in response to the specific credit risk of the borrowing firm. To address this identification challenge, I focus on the set of loans in my sample between a given bank and a given borrowing firm that are initiated in the same quarter, defined as a loan package. This involves regression models with third-party credit guarantee indicator as an explanatory variable and terms of lending such as loan rate as a dependent variable and controlling for bank–quarter–borrower fixed effects (or package fixed effects). The main assumption is that guaranteed and unguaranteed loans within one loan package have the same unobserved firm-specific risk and are affected by the same lender-specific characteristics and macroeconomic conditions at the time of loan initiation. I also control for the possible differences in the remaining loan characteristics such as loan amount, loan maturity, loan type, interest rate variability, collateral, and whether the loan is syndicated. Therefore, the empirical strategy must allow for the measurement of the marginal effect of a credit guarantee in a loan contract relatively accurately. The loan package fixed effects improve on previous identification strategies (Khwaja and Mian, 2008) as they jointly control for all unobservable time-varying firm and bank characteristics. About 16% of the sample loans are loans that belong to loan packages with at least two loans. A portion of these packages are packages that include at least one loan with a credit guarantee and at least one loan without a credit guarantee.[6]

Controlling for remaining differences in loan-specific characteristics, I find that within a loan package, a loan that is guaranteed by a third party receives on average a discount of 12 basis points (bps) relative to a loan that is not guaranteed. Considering the average interest rate of 3.291% for the loans used in the analysis, this is equivalent to a discount of about 4% in the cost of debt. The result that the existence of a credit guarantee significantly reduces the cost of borrowing for a firm is robust to the choice of controls and sample.

Having established the prevalence and effectiveness of third-party credit guarantees in debt financing, next I focus on the issue of substitutability of collateral and third-party credit guarantees by conducting two separate groups of tests. In the first group, I continue to use tests with package fixed effects by including collateral and guarantee indicators and their interaction term as explanatory variables. The analysis provides a number of results. First, I find that both collateral and guarantee reduce loan rate significantly. Second, I find that my earlier finding that credit guarantee reduces loan rate is not explained by the subsample of loans that are both guaranteed and collateralized. Third, not only do collateral and guarantees overlap in their effect on mitigating risk, but also each have their own specific effect on risk. Moreover, as expected, when I focus on loans that are uncollateralized, I find that the effect of a credit guarantee on cost of debt is even higher, about 25 bps and equivalent to 7.60% discount in cost of debt.

I explore the substitutability of collateral and third-party guarantees further in another group of tests by investigating how borrowing firms adjust their use of collateral versus credit guarantee in response to shocks to their collateral value. The data provide information on the general type of the collateral used by borrowing firms as well as the geographical location of the borrowing firms. Similar to Luck and Santos (2019), I focus on real estate collaterals because I can estimate the change in the value of real estate assets based on

6 A discussion of sample representativeness and extendibility is provided later in the paper.

geographical locations by using publicly available data. Controlling for firm-level and loan-level characteristics as well as bank–quarter–industry fixed effects. I find that borrowers that have experienced a positive shock to their real estate assets, defined as being from counties in the top quartile of annual real estate price growth, are 3.1 percentage points (significant at the 1% level) more likely to pledge real estate assets as collateral in their new borrowings. (As a point of comparison, about 9.3% of loans have real estate collateral.) Importantly, this likelihood is increased by 1.1 percentage points for firms that are smaller, defined as firms in the bottom quartile of firm size. Moreover, I find that small firms that experience a negative shock to their real estate assets—that is, firms from counties in the bottom quartile of annual real estate price growth—are 1.3 percentage points less likely to use real estate assets as collateral in their new borrowings. Moreover, I find that a negative shock to the real estate assets significantly increases the likelihood that a small firm uses a third-party credit guarantee (by 5.2%, significant at the 1% level). In other words, for a small firm, a negative shock to asset values significantly decreases the likelihood of pledging assets as collateral and significantly increases the likelihood of using a third-party credit guarantee to obtain new credit. This is particularly important considering that small firms play an important role in driving local economic growth and employment.

Next, I examine the mechanism through which a credit guarantee affects the interest rate charged by a lending bank. Controlling for a borrowing firm's risk (probability of default), a bank's expected loss of a loan is related to the loan's loss given default (LGD) and the bank's exposure at loan's default (ED) (Altman, 2011; BIS, 2017). A loan with a credit guarantee is expected to have a higher recovery rate (lower LGD) than a loan without a credit guarantee. To that end, I examine the effect of a credit guarantee on a loan LGD, as calculated by the lending bank, on the sub-sample of loan packages in the data with non-missing LGD estimates.[7] I find that within a loan package, a loan with a credit guarantee has a significantly lower LGD that is 81.6 bps lower than a loan without a credit guarantee (i.e., a reduction of 2.23% considering an average LGD of 36.25%). Moreover, taking advantage of quarterly loan status updates in the data, I find that within a loan package, guaranteed loans have 2.49 percentage points lower past-due rates than loans that are not guaranteed.[8] Therefore, the results confirm that guaranteed loans not only have lower bank estimated risks but also perform better over time.

There are various situations in which a bank extends two loans to a borrowing firm at the same time, where one loan has a credit guarantee and the other does not. Some of these situations are supply driven and some are demand driven. In light of the recent increase in demand for corporate loans from institutional investors (Gande and Saunders, 2012; Beyhaghi and Ehsani, 2016), banks try to structure loan packages in a way that attracts a large number of loan buyers. Some loan buyers prefer buying loans with simpler structures. Other loan buyers look for loans with specific characteristics or with more safety measures. Because of this, it is not surprising that banks provide loans to the same borrower at the same time but with different features. There are also demand-driven factors that might produce loan packages that include guaranteed and unguaranteed loans. A typical case stems from a limited or partial guarantee provision by a guarantor where the guarantor

7 Banks are required to report loan LGDs only in the second half of the sample period.

8 Note that the average loan past-due rate in the sample is just 2.89%. Also note that these values provide lower-bound estimates on the effect of a credit guarantee on LGD and performance as they ignore the "spillover" effect of third-party credit guarantees on package loan risk and default.

guarantees less than 100% of a bank's exposure to a borrowing firm. In this case, a bank might split its exposure to the borrowing firm into two loans in which one loan is fully guaranteed by the guarantor and the other loan is unguaranteed. This practice facilitates more effective loan risk management, loan portfolio management, loan pricing, and contract enforcement for banks.[9]

Further, I examine how the characteristics of the "source" of a credit guarantee—that is, the guarantor—might affect loan pricing. The source of a credit guarantee is important because a guarantor has its own credit risk, and the fact that a loan is guaranteed by another legal entity does not necessarily mean that the risk of borrowing firm non-payment is significantly reduced. The details in the data allow me to distinguish between three common types of guarantors: (1) a corporate guarantor, which is typically the borrower's parent or a group member; (2) a personal guarantor, which is an individual such as the business owner or a manager; and (3) an agency guarantor, which is a US government agency such as Small Business Administration or the Export–Import Bank of the USA. To strengthen identification, I further restrict the sample to loan packages where all loans in the package are uncollateralized. This practice also allows me to limit the sample size to a manageable number of loans for manual checking (about 5,000 loans). For this subsample, I manually check the identifying information of each guarantor, including name and contact information, to ensure that each guarantor is properly allocated to one of the above types.

Among the three types of guarantors, an agency guarantor has the highest credibility. An agency guaranteed loan is effectively backed by the US government, which is considered nearly riskless. Moreover, credit guarantee programs provided by the US government are standard, transparent, and have a long history of practice within the USA. The agency guarantee is, however, a special case in the corporate loan universe.[10] The most common forms of credit guarantees are corporate guarantees and personal guarantees. The advantage of these two types of guarantees is that since the parent company, in the case of a corporate guarantee, or the manager or the owner of the firm, in case of a personal guarantee, is directly involved in the governance of the borrowing firm, the lending bank can rely on the guarantor's additional incentive in repaying the loan on time. However, unlike agency guarantors, which are relatively riskless, these guarantors have credit risks of their own. An

9 A direct example is credit guarantees provided by US government agencies such as the Small Business Administration or the Export-Import Bank of the USA. These guarantors avoid guaranteeing 100% of bank exposures so that banks, which thus incur credit risk, have an incentive to monitor their contracts. Another example is the limited guarantee provided by a parent company, where the parent company does not have the ability or the will to guarantee all of the bank exposure. Based on law, if after giving effect to the guarantee the guarantor is "insolvent," then the requirements for a "fraudulent conveyance" or "fraudulent transfer" will have been established. To that end, the credit agreement is vulnerable to attack under federal and state insolvency laws. Put differently, it is illegal for a guarantor to undertake an obligation under a credit agreement that makes the guarantor's liabilities (including contingent liabilities) exceed its assets (Taylor and Sansone, 2006, pp. 328–332).

10 In 2020 (after the sample period), the US government initiated the first phase of an unprecedented credit guarantee scheme, the Paycheck Protection Program, under which the government acts as the third-party guarantor on $525 billion worth of new corporate loans that US financial institutions have extended to US enterprises. These loans have no collateral requirements. This program has dramatically increased the share of agency guarantee that is mentioned in this study.

advantage of a corporate guarantee over a personal guarantee is that a corporate guarantor generally has a more diverse asset portfolio, is more resourceful, and bears less credit risk. In contrast, a personal guarantor's wealth is generally tied to the borrowing firm's assets, making it unclear the extent to which the personal guarantor can repay a loan if the borrowing firm failed to.[11]

Consistent with the notion that different types of guarantors have different credit risks, I find that an agency-guaranteed loan receives on average a discount of 250 bps (a discount of 76%) relative to a loan that is not guaranteed, after controlling for lender and borrower time varying unobservable characteristics and the remaining differences in contract characteristics.[12] Considering that the average interest rate is 3.291%, this means that a credit guarantee provided by a US government agency reduces the level of interest rates to slightly less than 1%, which is close to the risk-free rate during the sample period. This finding provides an external validity of the empirical methodology: as expected, a government-guaranteed loan is nearly riskless and the lending bank expects a rate of return close to the risk-free rate. Moreover, a corporate guarantee generates a discount of 28.7 bps, which is equivalent to about 9% of the cost of debt for an average borrowing firm. I do not find any significant price effects for personal credit guarantees within a loan package for the loans in my sample (loans with $1 million or above in commitments).[13]

Finally, the main identification strategy in this study is based on using loan package fixed effects. To that end, for my primary analysis, I use the sample of borrowers with multiple loans initiated around the same time. I conduct further analysis to investigate how the specific sample selection affects my conclusions. If the sample of firms that have both guaranteed and unguaranteed loans (main sample) are riskier than the sample of firms that have only guaranteed loans, then the spread as estimated above is an overestimate of how much a guarantee affects loan rate. To investigate if this is the case, I divide the borrowers of guaranteed loans in the initial sample into two groups: (1) firms with only guaranteed loans and (2) firms with both guaranteed and unguaranteed loans. The results of both univariate and multivariate analyses reveal that firms with both guaranteed and unguaranteed loans are slightly less risky than firms with only guaranteed loans. Therefore, the results in my study do not overestimate the effect of a credit guarantee. Additionally, among the guaranteed loans, I do not find that being treated (being selected for the main sample) is significantly related to loan rates. These results are robust to different specifications and to including controls for firm characteristics, loan characteristics, and bank–quarter–industry

11 Note that the sample of loans for this study constitutes loans that are $1 million and above. For loans less than $1 million (small business loans), personal guarantees are much more effective than for larger loans, as the personal assets of the guarantor outside the ownership in the borrowing firm, such as the guarantor's real estate property, can provide sufficient backup for banks to recover their potential loss from a loan default.

12 As a comparison, Lelarge, Sraer, and Thesmar (2010) using data on French government's SBA program during 1989–2000 to estimate the discount in interest rate received by French firms under the program to be as large as 85%. They use a parametric selection model to estimate the magnitude of this discount.

13 As mentioned above, a reason that I do not find any significant result for personal guarantees might be that the data exclude small business loans. Prior studies show that personal guarantees are frequently used by banks as a mechanism to reduce the risk of small business loans (Berger and Udell, 1998; Rice and Strahan, 2010; Chaney, Sraer, and Thesmar, 2012; Schmalz, Sraer, and Thesmar, 2017).

fixed effects. In sum, I do not find evidence of selection issues or that loan rate effect of credit guarantee is overestimated in this study.

This study contributes to the literature in a number of ways. First, it shows the widespread use of third-party credit guarantees in the current corporate loan market. Second, it estimates the magnitude by which the borrowing firm and lending bank are affected across various dimensions: loan pricing, loan estimated LGD, and loan performance. Third, it discusses the structure of third-party credit guarantees in bank lending by analyzing various types of third-party credit guarantors and their impact on loan contracting. Fourth, it provides evidence that many firms, especially smaller firms, rely on third-party credit guarantee instead of or in addition to pledging their assets as collateral to obtain credit. The latter implies that the credit guarantee channel can mitigate the negative impact of the asset value channel for borrowing firms, specifically when firms do not have sufficient assets to pledge as collateral or when they face negative shocks to their asset values.

The remainder of this study is organized as follows. Section 2 describes the institutional background. Section 3 details the data and variables. Section 4 discusses the credit guarantee determinants and explains the empirical methodology. Section 5 provides the results. Section 6 concludes.

## 2. Institutional Background

### 2.1 Credit Guarantee Overview

Throughout this paper, a credit guarantee is an “explicit” guarantee, which is a legally binding commitment of the guarantor to pay an amount to the lender in case the borrowing firm defaults under its obligations to the lender.

There are other forms of guarantees that are used in the lending industry. These forms are known as implicit (or soft) guarantees and are generally provided by the borrowing firm’s parent company in forms of (1) comfort letters or letters of intent, which are basically a declaration rather than a legally binding commitment whereby the guarantor declares it will refrain from taking actions that would jeopardize its subsidiary’s financial stability, or in the form of (2) a keep-well agreement, which is a declaration by the parent company that it will provide its subsidiary with additional capital to prevent it from defaulting. Comfort letters, letters of intent, and keep-well agreements are used particularly when the borrowing firm is a subsidiary of a larger firm. According to PwC (2013), these implicit guarantees generally lack legal enforceability.

According to Gudger (1998), lenders prefer an explicit third-party credit guarantee over physical collateral since an explicit credit guarantee is easier to enforce via the legal system and is not subject to the problem presented by physical collateral such as its maintenance in good condition, verification of its value, and safekeeping. Prior empirical studies provide evidence on how banks reassess the value of outstanding collateral for their loans following legal changes (Cerqueiro, Ongena, and Roszbach, 2016), changes in collateral redeployability (Benmelech and Bergman, 2009), and changes in asset prices such as housing (Luck and Santos, 2019).

Despite some advantages over collateral, a third-party credit guarantee carries other risks: (1) a credit guarantor has credit risk of its own, and (2) even when the guarantor’s credit risk is low, sometimes the guarantee agreement cannot be enforced. The reason is that contracts are inherently incomplete (Hart and Moore, 1988; Dewatripont and Tirole, 1994) and it is not feasible for banks to stipulate guarantors’ obligations in all future-state contingencies. According to Moody's (2010), even in the case of explicit credit guarantees, depending on how the guarantee is drafted, the guarantor may be able to legally raise a wide array of defenses to its liability. For example, Moody's asserts that courts recognize a number of defenses specific to guarantees, which can allow a guarantor to materially delay or completely evade its guarantee obligation. These defenses may relate directly to the terms of the guarantee (e.g., failure to explicitly waive specific defenses) or to the terms of the loan in general (being unfair). In the latter case, the guarantor may attempt to limit or avoid liability for the borrowing firm. The new Basel instructions also recognize the legal risk associated with credit guarantees, especially when the borrowing firm is located in a country with weak legal system. BIS (2017) specifically instructs banks to employ robust procedures and processes to control the "residual risks" that are accompanied with credit guarantees including legal, operational, liquidity, and market risks.

Banks' exposure to guarantee risk, which is caused by incomplete contracting or contract enforcement (risk item #2 above) in the sample of loans in this study, is probably less severe than banks' exposure to guarantee risk as a result of the guarantor's own credit risk (risk item #1 above). The reason is that the loans in the sample are recent loans belonging to major US banks with a long history of corporate loan contracting. Moreover, the majority of loans are between US banks and US borrowers under the laws of the USA.[14] Additionally, in the data instructions, a full guarantee is defined as the existence of an "explicit recourse for full repayment of the credit obligation by a single guarantor."[15] A partial guarantee is defined as "explicit recourse for repayment of a portion of the credit obligation." According to the instructions, the guarantor is the legal entity that provides secondary support for repayment, whereas the borrowing firm (obligor) is the legal entity that provides the primary source of repayment.

In the next subsection, I analyze how guarantors' credit risk and type are relevant to the lending bank and the borrowing firm.

## 2.2 Types of Credit Guarantors

In this study, I assign credit guarantors to three general types: (1) corporate guarantors, (2) personal guarantors, and (3) US government agency guarantors.

The most well-known type of third-party credit guarantee is a "corporate guarantee." In general, corporate guarantors are firms with better cash flow, better asset quality, and an established ability to repay loans. The dominant form of corporate guarantee is the guarantee provided by a borrowing firm's parent company or a major group member.[16] Corporate guarantees are common for a number of reasons: Not only does the guarantor constitute a better and more diverse credit profile than the borrowing firm and is equipped with

14 For loans to foreign borrowing firms, I control for country-specific factors that might affect the value of a credit guarantee through controlling for time-varying country fixed effects.

15 Page 212, Field No. 44 of data instructions, available at https://www.federalreserve.gov/reportforms/forms/FR_Y-14Q20181231_i.pdf.

16 Corporate guarantees can be in the form of a downstream guarantee, where a parent company guarantees an obligation of its subsidiary. Other forms of corporate guarantees include cross-stream guarantees, where one subsidiary with better financial stand guarantees an obligation of another subsidiary, and upstream guarantees, where a major subsidiary guarantees the obligation of the parent. The latter case happens when the only asset of the parent company is stock ownership of a subsidiary.

additional assets to secure the repayment of the loan, but the guarantor also has sufficient information about the borrower and through its relationship with the borrowing firm, such as its stock ownership or trading relationship, directly benefits from the loan proceeds. As a result, the bank can rely on the extra monitoring of the corporate guarantor on the borrowing firm's risk taking. For the purpose of this study, I carefully read the identifying information of the borrowing firm and the guarantor in the main loan sample, including their names and geographical information, and I use Bloomberg.com and Google.com to identify whether the borrowing firm and the credit guarantor belong to the same group. I find that parent guarantee (downstream guarantee) is the most common form of credit guarantee.

Another type of credit guarantee that is used mostly for small business borrowers is a "personal guarantee," where the credit guarantee is provided by the business owner or manager. A personal guarantee reduces the risk for the lending bank because it gives the bank the right to pursue the owner's personal assets if the borrowing firm fails to repay the debt. The extent to which a bank can really rely on personal guarantees is not clear, as most of the time it is the case that the person who provides the guarantee does not have a diversified portfolio. One reason is that ownership in the firm constitutes the majority of the owner's wealth.[17] The bank, however, counts on the additional incentive that a personal guarantee might create for the managers to protect the bank's interests. If the credit guarantor is an individual, banks are asked not to disclose the specific name of the guarantor in FR Y-14Q filings; instead, they have to substitute the credit guarantor's name with the word "Individual."

The next group of guarantors includes various US government agencies whose mandate is to provide support for certain US businesses. Under this scheme, banks receive the credit guarantee from a government agency and make loans to certain firms. I call these guarantors "agency guarantors." According to Bachas, Kim, and Yannelis (2019), the justification for the US government to intervene in the credit market is that in the presence of information frictions, these guarantees may increase aggregate welfare by decreasing lenders' risk and encouraging lending. A well-known example of such a guarantee that appears in the data is the guarantees provided by the US Small Business Administration to support small businesses. Another example that also appears in the data is the guarantees provided by the Export–Import Bank of the USA (EXIM), whose mandate is to promote the export of American goods and services. Through this channel, an international buyer receives a loan from a US bank that is guaranteed by EXIM to finance import from a US firm. Other US government agencies that provide credit guarantees include but are not limited to the Overseas Private Investment Corporation, the Commodity Credit Corporation, and the United States Department of Agriculture Guaranteed Loan Program. An advantage of a US government-agency guarantee for banks is that since the guarantor is effectively the US government, the guarantor is considered riskless.[18]

17 Note that corporate loans that are the subject of this study do not include loans less than $1 million, which are typically loans to start-ups and generally small businesses. For these types of firms, an owner's personal wealth, specifically the owner's real estate property, is likely sufficient to cover most of a loan repayment.

18 To compare agency credit guarantee schemes across developed and developing countries see Beck, Klapper, and Mendoza (2008). Moreover, Wilcox and Yasuda (2019) discuss the unintended consequences of supplying large amounts of credit guarantees by the government on bank risk taking using Japan's Special Credit Guarantee Program in the 1990s.

Although all types of third-party credit guarantees mitigate concerns about a borrowing firm not being able to repay its loan, I expect that banks put more value on agency guarantees and less value on personal guarantees on mitigating the risk of loans in my sample.

## 3. Data Construction and Sample Descriptive Statistics

### 3.1 Federal Reserve's Y-14Q Schedule H.1

The data used in this study come from FR Y-14Q schedules that are collected quarterly from bank holding companies by the Federal Reserve. Following the implementation of the Dodd–Frank Wall Street Reform and Consumer Protection Act of 2010, large bank holding companies with total consolidated assets of $50 billion or more (about thirty-five bank holding companies) are required to disclose their portfolios to the Federal Reserve.[19] Particularly, I focus on the FR Y-14Q Schedule H.1, which collects details of corporate loans on banks' balance sheet. The data collection has two sections: (1) loan, obligor, and guarantor description section and (2) obligor financial data section. Both sections are completed at loan-level detail.

The collection of FR Y-14Q starts in the fall of 2011. As of June 2012, banks are required to disclose the details of all their corporate loans that are $1 million or more in commitment exposure (constituting 97% of bank corporate exposures), including information on credit guarantees and guarantors' identities.[20] As such, I start with the sample of all C&I loans to domestic and foreign obligors that are originated during the 2012–18 period.

I restrict the sample to all loans with a valid "guarantor flag." The guarantor flag is an indicator variable in the data that receives a value of 1 if the loan is fully guaranteed by a single guarantor other than a US government agency, a value of 2 if the loan is partially guaranteed by any guarantor, a value of 3 if the loan is fully guaranteed by a US government agency, and a value of 4 if the loan is not guaranteed. A loan is fully (partially) guaranteed only if there is explicit recourse for full (partial) repayment of the credit obligation by a guarantor.

Further, I restrict the sample to US dollar loans with a valid borrower identity excluding loans to borrowers identified as individuals.[21] I also exclude all unutilized commitments from the sample because banks report interest rates only when they receive interest payments on their loans.[22] Moreover, I exclude all loans that are contractually subordinated or their seniority status is not clear. As each loan appears in the data for multiple quarters, I keep only the first appearance of the loan in the data where the loan is utilized.

Further, to remove uncommon loans or loans with irregular features, I impose the following data restrictions: I remove loans that are neither a term loan nor a line of credit (e.g., fronting exposures and commitment to commit) and loans extended for purposes

19 Some bank holding companies only appear in the data for a relatively short time; others might not have a large C&I portfolio. In this study, I focus on bank holding companies that have at least 1,000 loans after the data cleaning process, which I elaborate on in the data section. The final sample includes loans belonging to the largest thirty bank holding companies.

20 The FR Y-14Q instruction that is published in June 2012 can be found at https://www.federalreserve.gov/reportforms/forms/FR_Y-14Q20120630_f.zip.

21 For some small borrowing firms, an individual, such as the manager or the founder, is considered the recipient of the credit.

22 Banks are required to report an interest rate of zero if the credit facility is unutilized.

other than capital expenditure, working capital, general corporate purposes, debt refinancing, new product development, and bridge financing (e.g., loans issued for asset securitization financing, commercial paper backup, mortgage warehousing, etc., are removed). Further loans whose interest rates are neither fixed nor floating are dropped from the sample (e.g., loans with mixed or unknown interest rate variability). Last, loans that are misreported, such as loans with negative amounts or interest rates higher than 100%, are removed. After imposing these restrictions, a sample of 148,590 unique loans remains.

### 3.2 Descriptive Statistics for the Total Sample

Table I presents descriptive statistics for the variables used in the regression analysis. A separate table for the subsample of loans used in loan package fixed effects analysis will be provided in the paper as well. All variables are described in detail in Appendix A. About 46% of sample loans and about 42% of sample loans are guaranteed and fully guaranteed by third-party guarantors, respectively. The average sample loan has an interest rate of 3.447% (median 3.250%). The average loan is for about $11 million with a time maturity of 46 months (less than 4 years). About 76% of loans are collateralized. Of the total sample, about 28% of loans are fixed interest rate loans, with the rest being floating rate loans and about 37% of loans are term loans with the rest being revolving credit. Moreover, about 27% of loans are considered syndicated loans.

Banks' estimated LGD ratio or LGD for an average loan is about 36% (median about 38%). About 3% of loans are past due 30 days or more during their life-time.[23] The median and the mean bank internal credit ratings assigned to a borrowing firm in the sample is about 6, which is equivalent to an S&P rating of BB. Banks are required to disclose the internal credit rating they assign to a borrowing firm and the corresponding S&P rating in FR Y-14Q filings.[24] As explained in Appendix A, the maximum possible internal credit rating is 10, which is equivalent to an S&P rating of AAA. The internal credit rating also includes the "soft" information that the bank has regarding the borrowing firm.

As noted by Brown, Gustafson, and Ivanov (2017), unlike the sample of firms in studies using COMPUSTAT and survey data, firms covered in the FR Y-14Q include all types of borrowing firms and are dominated by small private firms. Over 87% of loans are loans to US borrowing firms. The median book value of total assets is approximately US$75 million (log(assets) = 18.138). On average, the borrowing firms in our sample have a leverage ratio of 25.4% (defined as long-term debt-to-assets). They have a profitability ratio (net income to sales) of 5.2%, a liquidity ratio (current ratio) of 1.9, and a tangibility ratio (tangible assets to total assets) of 85%.

## 4. Empirical Methods

Having established the wide-spread use of third-party credit guarantees in the US corporate loan market, in this section, I describe the empirical strategy to identify the relation

23 This finding is consistent with other studies using post-2012 data, indicating that 2012–19 has been a relatively benign period in terms of loan defaults.

24 If an internal rating is mapped to a range of S&P ratings—for example, B–BB—instead of one S&P rating, I use the lowest S&P rating in the range as the borrower's credit rating (e.g., B from the B–BB range). The regression results stay qualitatively the same if I use the highest corresponding S&P rating or the median corresponding S&P rating.

**Table I.** Summary statistics

The table summarizes the loan-level data from Federal Reserve Y-14Q reporting forms. The sample period is from 2012 to 2018. The sample covers USD commercial and industrial loans with valid information originated during the sample period by large US bank holding companies (148,590 loans). All variables are defined in Appendix A.

Summary statistics at loan level

| | Mean | SD | 10% | Median | 90% |
|---|---|---|---|---|---|
| Credit guarantee | 0.464 | 0.499 | 0.000 | 0.000 | 1.000 |
| Fully guaranteed | 0.417 | 0.493 | 0.000 | 0.000 | 1.000 |
| Interest rate (prc) | 3.447 | 1.621 | 1.669 | 3.250 | 5.329 |
| Amount (million USD) | 10.718 | 36.368 | 0.488 | 2.556 | 25.222 |
| Log(amount) | 14.918 | 1.594 | 13.098 | 14.754 | 17.043 |
| Maturity (months) | 45.839 | 36.330 | 11.000 | 49.000 | 83.000 |
| Log(maturity) | 3.516 | 0.879 | 2.398 | 3.892 | 4.419 |
| Collateral | 0.762 | 0.426 | 0.000 | 1.000 | 1.000 |
| Fixed interest rate | 0.283 | 0.450 | 0.000 | 0.000 | 1.000 |
| Term loan | 0.368 | 0.482 | 0.000 | 0.000 | 1.000 |
| Syndicated loan | 0.274 | 0.446 | 0.000 | 0.000 | 1.000 |
| LGD (prc) | 36.217 | 14.026 | 16.140 | 37.740 | 50.000 |
| Past due (prc) | 2.895 | 16.767 | 0.000 | 0.000 | 0.000 |
| Domestic | 0.875 | 0.331 | 0.000 | 1.000 | 1.000 |
| Internal credit rating | 6.105 | 1.034 | 5.000 | 6.000 | 7.000 |
| Firm size (log(assets)) | 18.138 | 2.797 | 15.107 | 17.729 | 22.112 |
| Leverage (long-term debt/assets) | 0.254 | 0.250 | 0.000 | 0.199 | 0.580 |
| Profitability (net income/sales) | 0.052 | 0.131 | –0.013 | 0.036 | 0.158 |
| Liquidity (current ratio) | 1.896 | 2.023 | 0.624 | 1.379 | 3.271 |
| Tangibility (tangible assets/assets) | 0.851 | 0.225 | 0.479 | 0.975 | 1.000 |

between a credit guarantee and measures of loan pricing, LGD, and loan performance. Moreover, for the subset of loans that I identify the source of credit guarantee, I present the empirical strategy to identify and compare the effects of corporate, personal, and US government-agency guarantees on the variables of interest. The results of the empirical analysis are provided in Section 5.

## 4.1 Methodology: Loan Package Fixed Effects

The purpose of this subsection is to provide a methodology to accurately estimate the marginal effect of a credit guarantee on loan pricing, estimates of loan risk, and loan performance.

In order to remove confounding risk factors at the firm level that affect the existence of a credit guarantee and the subsequent loan pricing, I use a setup where a loan rate is regressed on the credit guarantee indicator by considering bank–quarter–borrower[25] fixed

25 Note that the quarter fixed effect here means that all loan observations within the same quarter group are originated in the same quarter and also are reported the same quarter, in case the quarter of origination is different from the quarter of reporting in the data.

effects. The idea is to take advantage of borrowing firms in the data that receive more than one loan from a bank in a given quarter. Particularly, I am interested in "loan packages," defined as loans between a borrowing firm and a bank issued in the same quarter, that include at least one loan with a credit guarantee and at least one loan without a credit guarantee. As mentioned in Section 1, the main assumption in this study is that different loans within one loan package have the same unobserved firm-specific risk and are affected by the same lender-specific characteristics and macroeconomic conditions at the time of loan initiation.

To strengthen identification, I control for possible heterogeneity in loan characteristics across a firm's loans that are initiated at the same time. As mentioned in the data section, all loans that are contractually subordinated or for which the seniority status is missing are already excluded from the study. The loan-level control variables considered include loan amount, loan maturity, collateral, loan rate variability, loan type, and syndicated status.

After preparing the sample, the next step is estimation. My baseline approach is to estimate the following multivariate linear regression model via ordinary least squares (OLS):

$$Y_{ij} = \alpha_i + \alpha_j + \beta \text{CreditGuarantee}_j + \gamma \text{LoanControls}_j + \epsilon_{ij}, \tag{1}$$

where $Y_{ij}$ represents an outcome of interest related to loan $j$ in loan package $i$, such as loan terms including loan rate, loan amount, loan maturity, lending bank estimated LGD, and loan future performance. $\text{CreditGuarantee}_j$ is an indicator variable equal to 1 if loan $j$ is guaranteed by a third party. $\text{LoanControls}_j$ includes a set of loan characteristics that are controlled.

To strengthen identification further, I restrict the sample to loan packages where none of the loans in the package are collateralized and repeat the analysis. Although I control for collateral in all specifications, this additional step ensures that the results in my study are not driven by the subsample of loans that are both collateralized and guaranteed.[26]

Later on in the paper, I estimate by OLS regression models of the form

$$Y_{ij} = \alpha_i + \alpha_j + \sum_{k=0}^{4} \beta_k \text{CreditGuaranteeType}_{jk} + \gamma \text{LoanControls}_j + \epsilon_{ij}, \tag{2}$$

where credit guarantee type represents the type of third-party credit guarantee out of three possible types: corporate, personal, and US government agency.

In order to determine the type of credit guarantee, I carefully study the details of each sample loan in the data including the identities of the borrowing firms and guarantors and manually assign each guarantor to one of the above groups. For personal guarantors, banks are asked to report the word "Individual" instead of the actual name of the guarantor.[27] For the remaining guarantor names, I conduct a Bloomberg search and a Google search to see whether the obligor belongs to the same corporate group as the guarantor, and if not, whether the guarantor is a government agency. Moreover, banks are required to report in a separate field in the data if a loan is fully guaranteed by a US government entity, so I double-check the guarantors I consider as agency guarantors and investigate any discrepancy.

26 A separate analysis of the relation between collateral and third-party guarantees is provided later in the paper.

27 Banks are not required to report the guarantor name if the guarantor is a natural person. However, instead of guarantor's name, they should report "Individual."

## 5. Results

### 5.1 Effect on Loan Rates

If third-party credit guarantees are used frequently to mitigate risk, to what extent do they affect the cost of borrowing? As mentioned earlier, the magnitude of discount in the cost of debt has important implications for the borrowing firm. The negative relation between cost of debt and future investments, for example, has been discussed extensively in the literature (Philippon, 2009; Gilchrist and Zakrajšek, 2012; Frank and Shen, 2016).

The idea is to consider bank–quarter–borrower (loan package) fixed effects. Adding loan package fixed effects automatically drops some loan observations from the analysis. These are loans for which there is no other loan in the sample, where the borrowing firm, the lending bank, and the quarter of originations are the same. The final loan sample effectively used in the analysis includes about 17% of the loans in the original sample. To ensure that the characteristics of these loans are comparable to the loans in the initial analysis, I report the summary statistics of the new sample (I call it the main sample) in Appendix B and provide additional detailed analysis later on in this section.

As expected, loans in the main sample are slightly larger than loans in the initial sample (mean loan amount of $13.643 million versus $10.718 million) and belong to larger borrowing firms (a log(assets) of 19.135 versus a log(assets) of 18.138). Also, the percentage of loans to domestic (US) borrowers is higher in the main sample relative to the initial sample (90.5% versus 87.5%). Moreover, loans in the main sample on average have slightly lower interest rates relative to loans in the initial sample (a mean of 3.291% versus 3.447%). They have longer loan maturities (53 months versus 46 months), are more likely to be term loans (47% versus 37%), and are more likely to be syndicated (44% versus 27%). The difference between main sample loans and initial loans across other loan characteristics, including percentage of being collateralized or being fixed rate, is relatively small. Moreover, loans in both samples, on average, have very close LGD ratios. Loans in the main sample have slightly lower future past-due ratios (2.592% versus 2.895%) in the initial sample. Together with having lower interest rate, one can conclude that loans in the main sample are slightly safer than loans in the initial sample. This means that the magnitude of discount in cost of borrowing that will be calculated in the following, using the identification strategy outlined in the previous section, might be an underestimation of the actual discount in the cost of borrowing. I discuss this further using multivariate analysis later on in this section.

Table II presents the results. The variable of interest in both columns of the table is the credit guarantee indicator.

In Column 1, which has only the third-party credit guarantee indicator as the explanatory variable, the point estimate for the indicator is negative (–0.093) and statistically significant at the 10% confidence level. The direction of this estimate is consistent with our expectation: the existence of a credit guarantee negatively affects a loan rate. Further, adding additional controls strengthens the finding. The results in Column 2 of Table II suggest that controlling for remaining differences among loans, an existence of a third-party credit guarantee reduces a loan spread by 119 bps (significant at the 5% level). Considering that the average interest rate is 3.291%, as shown in Appendix B, this is equivalent to a 3.62% discount on cost of debt.

The price effect of a credit guarantee should be particularly pronounced when the loan risk is higher. To that end, in Table III, I repeat the analysis in Column 2 of Table II, but

**Table II.** Third-party credit guarantee and loan rates

This table shows the effect of third-party credit guarantee for loan pricing in a set-up that lender, borrower, and time fixed effects are jointly controlled. The unit of observation in each regression is a loan. The dependent variable is loan rate (%). The sample period is from 2012 to 2018. The source of data is Federal Reserve Y-14Q reporting forms. All columns include controls for loan package fixed effects. All variables are defined in Appendix A. Standard errors are clustered at the firm level. ***, **, and * denote 1%, 5%, and 10% statistical significance, respectively (*P*-values are reported in parentheses).

| | Interest rate | |
|---|---|---|
| | (1) | (2) |
| Credit guarantee | –0.093* | –0.119** |
| | (0.084) | (0.014) |
| Log(amount) | | –0.083*** |
| | | (0.000) |
| Log(maturity) | | 0.067*** |
| | | (0.000) |
| Collateral | | –0.263*** |
| | | (0.003) |
| Fixed interest rate | | 0.408*** |
| | | (0.000) |
| Term loan | | 0.105*** |
| | | (0.000) |
| Syndicated loan | | 0.035 |
| | | (0.812) |
| Bank–quarter–borrower–origination quarter FE | Yes | Yes |
| Observations | 24,573 | 23,520 |
| Adjusted $R^2$ | 0.86 | 0.86 |

this time I include the interaction of the third-party credit guarantee dummy with the indicator for loan collateralization. The results show that the price effect of a credit guarantee is more pronounced for loans that are not collateralized. The point estimates for the credit guarantee indicator in specifications with and without other loan controls are, respectively, –0.296 and –0.326, both statistically significant at the 1% confidence level.

The results in Table III provide other intuitions. First, the finding that third-party credit guarantees reduce cost of debt is not driven by the sample of loans that are also collateralized. Second, considering the coefficient of the interaction term, the results imply that the effect of a third-party credit guarantee and the effect of collateral partly overlap. The coefficients for the interaction term between the credit guarantor dummy and the collateral dummy under both specifications in Table III are positive and significant at the 5% level (about 0.26), implying that credit guarantee and collateral mitigate the impact of each other on the cost of debt (i.e., they are substitutes). A loan that is only guaranteed has an interest rate that is about 33 bps lower relative to an unguaranteed but otherwise similar loan (Specification 2). Similarly, a loan that is only collateralized has an interest rate that is about 34 bps lower than a similar but uncollateralized loan. However, the discount for a loan that is both guaranteed and collateralized is less than the sum of individual discounts,

**Table III.** Third-party credit guarantee versus collateral

This table shows the effect of third-party credit guarantee versus collateral for loan pricing in a set-up that lender, borrower, and time fixed effects are jointly controlled. The unit of observation in each regression is a loan. The dependent variable is loan rate. The main explanatory variables are credit guarantee indicator, collateral indicator, and the interaction term between these two indicators. The source of data is Federal Reserve Y-14Q reporting forms. All columns include controls for loan package fixed effects. All variables are defined in Appendix A. Standard errors are clustered at the firm level. ***, **, and * denote 1%, 5%, and 10% statistical significance, respectively (*P*-values are reported in parentheses).

| | Interest rate | |
|---|---|---|
| | (1) | (2) |
| Credit guarantee | –0.296*** | –0.326*** |
| | (0.002) | (0.000) |
| Log(amount) | | –0.083*** |
| | | (0.000) |
| Log(maturity) | | 0.067*** |
| | | (0.000) |
| Collateral | –0.339*** | –0.343*** |
| | (0.002) | (0.001) |
| Credit guarantee × collateralized | 0.262** | 0.264** |
| | (0.013) | (0.012) |
| Fixed interest rate | | 0.408*** |
| | | (0.000) |
| Term loan | | 0.105*** |
| | | (0.000) |
| Syndicated loan | | 0.039 |
| | | (0.792) |
| Bank–quarter–borrower–origination quarter FE | Yes | Yes |
| Observations | 24,573 | 23,520 |
| Adjusted $R^2$ | 0.86 | 0.86 |

about 42 bps (= 33 + 34 – 26). The overlap indicates that a third-party credit guarantee and collateral are partly substitutes. However, each has its own individual effect on the rate (they also complement each other). As mentioned in Section 2, third-party credit guarantees have a number of advantages over collateral. Credit guarantees are easier to enforce via the legal system and are not subject to the problem presented by collateral such as its maintenance in good condition, verification of its value and safekeeping (Gudger, 1998). On the other hand, third-party credit guarantees might have other disadvantages over collateral. Most importantly, a third-party guarantor has credit risk of its own and the fact that a loan is fully guaranteed by a third party does not mean that the loan will be certainly paid off.

To ensure that the results in Tables III and IV are not driven by unobserved loan characteristics, in Table IV, I repeat the main analysis used in Table II, but this time I limit the sample to only bi-lateral loans (Column 1), to only fixed-rate loans (Column 2), to only fully collateralized loans (Column 3), and to only uncollateralized loans (Column 4). The differing results are consistent with my prior findings. However, the results should be

**Table IV.** Third-party credit guarantee and loan rates: further tests

This table shows the effect of third-party credit guarantee for loan pricing in a set-up that lender, borrower, and time fixed effects are jointly controlled. Columns 1–4, respectively, show the results of analysis separately across bi-lateral loan facilities, fixed-rate loan facilities, fully collateralized loan facilities, and uncollateralized loan facilities. The unit of observation in each regression is a loan. The dependent variable is loan rate (%). The source of data is Federal Reserve Y-14Q reporting forms. All columns include controls for loan package fixed effects. All variables are defined in Appendix A. Standard errors are clustered at the firm level. ***, **, and * denote 1%, 5%, and 10% statistical significance, respectively (*P*-values are reported in parentheses).

| | Bi-lateral loans (1) | Fixed rate loans (2) | Fully collateralized loans (3) | Uncollateralized loans (4) |
|---|---|---|---|---|
| Credit guarantee | –0.081* | –0.188** | –0.078 | –0.250*** |
| | (0.092) | (0.014) | (0.149) | (0.007) |
| Log(amount) | –0.050*** | –0.005 | –0.092*** | –0.049*** |
| | (0.000) | (0.823) | (0.000) | (0.002) |
| Log(maturity) | 0.017 | –0.015 | 0.064*** | –0.004 |
| | (0.442) | (0.837) | (0.000) | (0.947) |
| Collateral | –0.091 | –0.261* | | |
| | (0.219) | (0.079) | | |
| Fixed interest rate | 0.540*** | | 0.435*** | 0.178** |
| | (0.000) | | (0.000) | (0.046) |
| Term loan | 0.133*** | 0.058 | 0.108*** | 0.095*** |
| | (0.000) | (0.569) | (0.000) | (0.005) |
| Syndicated loan | | 0.965* | –0.243* | 0.716** |
| | | (0.073) | (0.051) | (0.033) |
| Bank–quarter–borrower–origination quarter FE | Yes | Yes | Yes | Yes |
| Observations | 12,578 | 5,546 | 17,653 | 5,179 |
| Adjusted $R^2$ | 0.89 | 0.86 | 0.88 | 0.84 |

interpreted with caution as the sample size drops with these restrictions. The results when I distinguish between fully collateralized and uncollateralized loans indicate that third-party credit guarantees are particularly important when there is no collateral pledged with a loan. The results in Column 4 indicate that the existence of a third-party credit guarantee is associated with an average reduction of 25 bps in cost of debt or about 8% in cost of debt.

## 5.2 Effect on LGD and Loan Performance

If third-party credit guarantees make loans safer and if this is the reason banks are willing to provide a discount on cost of debt, then loans with a credit guarantee must have lower LGDs and better performance over time. In addition to providing LGD ratios, the quarterly updates on loan status in FR Y-14Q data enable me to track a loan over time and see if a loan is past due at any time during its life. A loan is identified as past due in FR Y-14Q if the loan is past due by at least 30 days.

**Table V.** Third-party credit guarantee, loan performance, and loan perceived risk

This table shows the effect of third-party credit guarantee on bank's estimated LGD ratio as well as loan future performance in a set-up that lender, borrower, and time fixed effects are jointly controlled. Loan future performance is measured by whether the loan will be past due by 30 days or more in any quarter after loan origination. The unit of observation in each regression is a loan. The source of data is Federal Reserve Y-14Q reporting forms. All columns include controls for loan package fixed effects. All variables are defined in Appendix A. Standard errors are clustered at the firm level. ***, **, and * denote 1%, 5%, and 10% statistical significance, respectively (*P*-values are reported in parentheses).

| | LGD (1) | Past-due (2) |
|---|---|---|
| Credit guarantee | –0.816* | –2.490* |
| | (0.092) | (0.057) |
| Log(amount) | –0.388*** | 0.301** |
| | (0.000) | (0.025) |
| Log(maturity) | –0.088 | –1.162** |
| | (0.615) | (0.015) |
| Collateral | –7.883*** | –0.424 |
| | (0.000) | (0.667) |
| Fixed interest rate | –0.498* | –0.895 |
| | (0.072) | (0.293) |
| Term loan | 1.486*** | –1.291*** |
| | (0.000) | (0.000) |
| Syndicated loan | –2.437*** | 2.604 |
| | (0.002) | (0.265) |
| Bank–quarter–borrower–origination quarter FE | Yes | Yes |
| Observations | 12,573 | 16,046 |
| Adjusted $R^2$ | 0.89 | 0.45 |

Within a loan package, the loan with a third-party credit guarantee is expected to have a higher recovery rate than the loan without a credit guarantee. Therefore, I expect that the lending bank assigns a lower LGD to the loan with a credit guarantee. Moreover, I expect that the loan with a third-party credit guarantee outperforms a loan without a third-party guarantee. Therefore, a guaranteed loan is less likely to be past due.

The results of LGD and loan performance analysis are presented in Table V. In Column 1, the ratio of LGD is regressed on credit guarantee indicator and a set of controls where time-varying borrower and bank fixed effects are controlled. The coefficient for loan LGD indicates that a third-party credit guarantee reduces LGD by about 0.82%, which is equivalent to 2.2% of the average LGD. One should note that this is an underestimation of effect on LGD assuming that the guarantee might have some spillover effect on the LGD of all loans in the same loan package.

More importantly, a similar analysis in Column 2, where past-due indicator is the dependent variable, reveals that a loan with a third-party credit guarantee is 2.490 percentage points less likely to be past due during its life relative to a similar loan without a credit guarantee. Considering that the average past-due rate during the sample period is 2.592%, this finding indicates that a credit guarantee reduces the likelihood of being nonperforming to a level close to zero during the sample period.

In summary, consistent with the notion that a third-party credit guarantee reduces risk, and as a result it is accompanied with a reduction in the cost of borrowing, I also find that loans with third-party guarantees also have lower LGD ratios and lower nonperforming rates.

## 5.3 Distinguishing between Corporate, Personal, and Agency Guarantors

To what extent does the effect of third-party credit guarantee depend on the party that provides the credit guarantee? In Section 2, I argue that a credit guarantor has a credit risk of its own, and therefore, the fact that a loan is guaranteed by a third party does not mean that the credit risk of the loan is completely eliminated. To that end, I expect a credit guarantee to be more effective when the credit guarantor is safer.

To examine, I continue with the sample of loan packages where all loans in the package are uncollateralized (similar to the sample used in the last column of Table IV). There are two reasons for choosing this subsample to measure the coefficient magnitude of each type of third-party credit guarantee: (1) removing collateralized loans strengthens the identification and (2) this practice also allows me to limit the sample size to a manageable number of loans for manual checking (about 5,000 loans). For this subsample, I check the identities of the guarantor as provided in the FR Y-14Q data, including guarantor names, along with other complementary information provided in FR Y-14Q, and assign guarantors into three groups as described in Sections 2 and 4. These groups include: corporate guarantors, personal guarantors, and US government-agency guarantors. I run the analysis for main loan terms (loan rate, loan amount, and loan maturity). The results are provided in Table VI.[28]

The results confirm that the source of credit guarantee is important. In Column 1 of Table VI, the coefficient for credit guarantee identifier, when I do not distinguish for the source of guarantee, is –25 bps (equivalent to about 7.6% discount in cost of debt). However, when I distinguish between the sources of guarantee in Column 2, the results indicate that a US government-agency guarantee has a significantly higher rate deduction effect on cost of debt than any other type. The coefficient for an agency guarantee is about 250 bps, which is equivalent to 75.9% discount in cost of debt (when compared with average loan rate of 3.291%). This discount is significant at the 1% level. This discount also implies that an agency guarantee makes a risky loan almost riskless from the lending bank's perspective (the rate is lowered to the level of the risk-free rate, i.e., the rate on government debt) as the loan is effectively guaranteed by the US government.[29]

28 Note that due to the drop in the sample size of loans with nonmissing LGD and firm performance and the infrequency of remaining loans in some of subgroups, the results for these variables of interest are not produced. However, the results at an aggregate level are provided in Table V.

29 A related study on the effect of government-backed credit guarantee programs on alleviating credit constrains is the study by Lelarge, Sraer, and Thesmar (2010). Lelarge, Sraer, and Thesmar (2010) study the French government's Small Business Administration program during the 1989–2000 period. Similar to the results in this paper, they find that obtaining a government-guaranteed loan significantly decreases the effective interest rate for firms. The estimated reduction in interest rate is 6 percentage points according to their matching/OLS estimates and up to 23 percentage points according to their parametric selection model. As a reference, the median interest rate

**Table VI.** Corporate, personal, and agency credit guarantees: The effect of the source of credit guarantee on loan pricing

This table shows the effect of third-party credit guarantee for loan pricing in a set-up that lender, borrower, and time fixed effects are jointly controlled. The sample is restricted to uncollateralized loans for which the identity of third-party guarantee (if any) is manually verified. The dependent variable in Columns 1 and 2 is loan rate (%). The dependent variable in Columns 3 and 4 is log(loan amount in USD). The dependent variable in Columns 5 and 6 is log(loan maturity in months). The unit of observation in each regression is a loan. The source of data is Federal Reserve Y-14Q reporting forms. All columns include controls for loan package fixed effects. All variables are defined in Appendix A. Standard errors are clustered at the firm level. ***, **, and * denote 1%, 5%, and 10% statistical significance, respectively (*P*-values are reported in parentheses).

| | Interest rate | | Amount | | Maturity | |
|---|---|---|---|---|---|---|
| | (1) | (2) | (3) | (4) | (5) | (6) |
| Credit guarantee | –0.250*** | | 0.070 | | 0.066 | |
| | (0.007) | | (0.737) | | (0.309) | |
| Credit guarantee—US Gov. Agency | | –2.502*** | | 2.955*** | | –0.225 |
| | | (0.000) | | (0.000) | | (0.107) |
| Credit guarantee—Corporate | | –0.287** | | 0.134 | | 0.091 |
| | | (0.010) | | (0.567) | | (0.191) |
| Credit guarantee—Personal | | –0.051 | | 2.445*** | | –1.866*** |
| | | (0.683) | | (0.000) | | (0.000) |
| Log(amount) | –0.049*** | –0.048*** | | | –0.004 | –0.002 |
| | (0.002) | (0.003) | | | (0.674) | (0.774) |
| Log(maturity) | –0.004 | –0.004 | –0.019 | –0.013 | | |
| | (0.947) | (0.941) | (0.674) | (0.774) | | |
| Fixed interest rate | 0.178** | 0.179** | –0.152 | –0.155 | 0.063 | 0.064 |
| | (0.046) | (0.046) | (0.202) | (0.193) | (0.298) | (0.292) |
| Term loan | 0.095*** | 0.096*** | 0.657*** | 0.659*** | 0.079*** | 0.076*** |
| | (0.005) | (0.005) | (0.000) | (0.000) | (0.000) | (0.000) |
| Syndicated loan | 0.716** | 0.755** | 0.176 | 0.123 | 0.319*** | 0.323*** |
| | (0.033) | (0.026) | (0.379) | (0.533) | (0.008) | (0.008) |
| Bank–quarter–borrower–origination quarter FE | Yes | Yes | Yes | Yes | Yes | Yes |
| Observations | 5,179 | 5,179 | 5,179 | 5,179 | 5,179 | 5,179 |
| Adjusted $R^2$ | 0.84 | 0.84 | 0.62 | 0.62 | 0.82 | 0.82 |

Moreover, the coefficient for corporate guarantee in Column 2 of Table VI implies that of the remaining types of credit guarantees, corporate guarantee is the most valuable. A corporate guarantee is associated with a discount in cost of debt that is on average about 28.7 bps (equivalent to 8.7% discount in cost of debt). As discussed before, a parent company is directly involved in the corporate governance of the borrowing firm. Also, a parent is 12% and the mean of this variable in their study is about 27%. Therefore, their estimates suggest a 22–85% decrease in cost of debt as a result of a government guarantee.

company directly benefits from the loan proceeds through its stock ownership of the borrowing firm. Therefore, a corporate guarantee provides a valuable support for bank lending.

The results also suggest that the point estimate for personal guarantee indicator is not significantly different from zero. Personal guarantees, although provided by an owner or a manager of a borrowing firm, have limited value in supporting large borrowings (loans with $1 million and above are the subject of this study) since the personal wealth of the guarantors is usually highly correlated with their investments in the borrowing firm, and managers' or shareholders' real estate ownership (major personal investments outside stock ownership) is unlikely to be sufficient for repayment of large loans. As such, if the borrowing firm defaults, the extent to which the bank can recover its losses from the personal guarantor is not clear.

Other important loan terms, namely loan amount and loan maturity, are considered in Columns 3–6 of Table VI. The results show that at an aggregate level, the effect of third-party credit guarantees on loan amount and loan maturity is relatively modest. However, the results in Column 4 indicate that loans with an agency guarantee and loans with a personal guarantee are significantly larger. Interestingly, loans with a personal guarantee generally have a shorter maturity than loans without a personal guarantee.[30] In sum, I conclude that the majority of the effect of third-party credit guarantee is reflected in the loan rate.

## 5.4 Sample Selection Process and Estimation of Loan Rate Discount

To what extent are the results based on the main sample extendable to the entire sample? If the sample of firms that have both guaranteed and unguaranteed loans are riskier than the sample of firms that have only guaranteed loans, then the spread as estimated using the main identification strategy is an overestimate of how much a guarantee affects loan rate. If selection is an issue, then the results shown in previous sections are the net effect of two channels: the selection channel and the guarantee channel. To investigate, I start with the initial data and provide a comparison across three subsamples of borrowing firms: (1) firms that have only guaranteed loans, (2) firms that have only unguaranteed loans, and (3) firms in the main sample, which include firms with multiple loans in a quarter where some loans are guaranteed and others are not.

In Table VII Panel A, these subsamples are compared based on the observable firm characteristics as well as banks' confidential internal credit rating that is assigned to each borrower. Columns 1–3 of Panel A report the mean values of rating and financials, respectively, for firms in subsamples 1, 2, and 3. Columns 4 and 5 of Panel A, respectively, report the difference between variable means in Columns 1 and 3, and the difference between variable means in Columns 2 and 3. The significance levels of a two-sample $t$-test for each difference are also reported.

The results show that an average firm across all subsamples has a borrowing rating of about 6, which is equivalent to an S&P rating of B. The results also show that firms with

30 Caution needs to be paid with respect to results for loan amount and loan maturity because as mentioned before, the amount and the length of bank exposure to be guaranteed is not only the bank's decision. It is also a decision that needs to be made by the guarantor as to how much of and for how long a bank's exposure to the borrowing firm will be guaranteed. So unlike loan rate, loan amount and loan maturity are not determined mainly by risk considerations.

**Table VII.** Likelihood of being in the main sample

This table investigates the firm-level differences across three subsamples: (1) firms that have only guaranteed loans, (2) firms that have only unguaranteed loans, and (3) firms in the main sample, which include firms with multiple loans in a quarter where some loans might be guaranteed and the others are not. The unit of observation is a firm. The sample covers firms that borrowed new C&I loans from banks during 2012–2018. Panel A provides a univariate analysis, and Panel B provides a multivariate analysis, in which the dependent variable is an indicator of being in the main sample. In the first two columns of Panel B, the sample includes subsamples 1 and 3. In the last two columns of Panel B, the sample includes subsamples 2 and 3. Standard errors are clustered at the firm level. ***, **, and * denote 1%, 5%, and 10% statistical significance, respectively (*P*-values are reported in parentheses).

Panel A: Likelihood of being in the main subsample—univariate analysis (firm level)

| | Sample mean | | | Difference with main sample | |
|---|---|---|---|---|---|
| | Only guaranteed (1) | Only unguaranteed (2) | Main sample (3) | Only guaranteed (4) | Only unguaranteed (5) |
| Internal credit rating | 6.065 | 6.106 | 6.110 | 0.001 | 0.066*** |
| Firm size (log(assets)) | 17.686 | 18.933 | 19.399 | –1.950*** | –0.591*** |
| Leverage (long-term debt/assets) | 0.248 | 0.257 | 0.325 | –0.093*** | –0.066*** |
| Profitability (net income/sales) | 0.048 | 0.056 | 0.054 | –0.003** | 0.004** |
| Liquidity (current ratio) | 1.929 | 2.073 | 1.663 | 0.182*** | 0.315*** |
| Tangibility (tangible assets/assets) | 0.863 | 0.841 | 0.776 | 0.090*** | 0.059*** |

Panel B: Likelihood of being in the main subsample—multivariate analysis (firm level)

| | Relative to only guaranteed | | Relative to only unguaranteed | |
|---|---|---|---|---|
| | (1) | (2) | (3) | (4) |
| Internal credit rating | 0.007*** | 0.007*** | –0.009*** | –0.009*** |
| | (0.000) | (0.000) | (0.000) | (0.000) |
| Firm size (log(assets)) | | 0.030*** | | 0.006*** |
| | | (0.000) | | (0.000) |
| Leverage (long-term debt/assets) | | 0.103*** | | 0.102*** |
| | | (0.000) | | (0.000) |
| Profitability (net income/sales) | | 0.064*** | | 0.038*** |
| | | (0.000) | | (0.003) |
| Liquidity (current ratio) | | –0.004*** | | –0.006*** |
| | | (0.000) | | (0.000) |
| Tangibility (tangible assets/assets) | | –0.054*** | | –0.089*** |
| | | (0.000) | | (0.000) |
| Bank–quarter–country–industry FE | Yes | Yes | Yes | Yes |
| Observations | 64,443 | 47,719 | 71,087 | 50,846 |
| Adjusted $R^2$ | 0.13 | 0.18 | 0.07 | 0.08 |

only unguaranteed loans are slightly better rated (less risky) than firms in the main sample, which is consistent with the discussion provided in Section 1 that a credit guarantee on a loan contract is a signal of the borrowing firm's unobserved riskiness. The difference in rating is, however, very small. The results of the univariate analysis do not show any significant difference between the sample of firms in the main sample and the sample of firms with only guaranteed loans. Looking at the firm financials, I find that, as expected, firms that do not have multiple loans in the same quarter are on average larger than firms in the other two subgroups. Moreover, firms in the main sample have higher leverage ratios (by less than 10%) relative to the other subsamples. They have slightly higher profit margins relative to only guaranteed and slightly lower profit margins relative to only unguaranteed. Moreover, they have lower current ratio and lower tangibility relative to the other subsamples.

The multivariate analysis as provided in Panel B of Table VII provides a more accurate account for the selection issue because it controls for the bank–quarter–country–industry fixed effects. In Panel B, an indicator for the inclusion in the main sample is the dependent variable. In Columns 1 and 2 of Panel B, the sample for analysis includes the subsample of firms with only guaranteed loans in addition to firms in the main subsample. In Columns 3 and 4 of Panel B, the sample includes the subsample of firms with only unguaranteed loans in addition to firms in the main subsample. Bank internal credit rating is the main explanatory variable in the regressions. I also use other firm characteristics as controls. The results show that relative to firms with only guaranteed loans, the firms in the main sample are slightly less risky. The results also show that relative to firms with only unguaranteed loans, the firms in the main sample are slightly more risky. The difference in riskiness as shown by bank internal rating is less than one-tenth of one rating level. When I include firm financials, the coefficient for bank internal credit rating remains the same, which indicates that a bank's confidential rating already includes the observable firm characteristics.

In sum, the results of Table VII indicate that since firms in the main sample are not riskier than firms with only guaranteed loans, the spreads I find in previous sections therefore do not overestimate the effect of a credit guarantee.

I also explore to what extent selection is an issue in driving my results—that is, affecting the spread between a guaranteed loan and an otherwise similar unguaranteed loan. My goal is to decompose the total effect of the guarantee in the selection channel and the pure guarantee channel to the extent that data permit. Table VIII presents the results. In Table VIII, I estimate loan-level regressions in which all the guaranteed loans in the initial sample, regardless of being in the main sample or not, are being used. Loan rate is the dependent variable. I use an indicator for "treatment" as the main explanatory variable, which takes a value of 1 if the guaranteed loan is in the main sample, and zero otherwise and control for all observable firm and loan characteristics as described in the other tables. If being selected into the sample is systematically related to a lower spread, then the selection channel explains a part of my earlier results. Otherwise, the results I find are predominantly explained by the pure guarantee channel. The results in Column 1 show that being in the treatment sample does not significantly relate to the loan rate. To ensure that this result is not driven by the sample of loans that are collateralized in addition to being guaranteed, in Column 2 I exclude all collateralized loans from the sample. The results stay qualitatively the same. In Column 3, I add an interaction term between the treatment indicator and bank internal rating. The purpose of this exercise is to see if being selected into the sample has different effects for different rating categories. I do not find any evidence supporting this

**Table VIII.** In-sample observations and loan rates

This table investigates the effect of selection into the main subsample on driving the results in the study's main analyses. The unit of observation is a loan. The sample covers all new C&I loans issued by banks that are guaranteed during 2012–2018. The dependent variable in all regressions is a loan rate. In Column 1, all guaranteed loans are used. In Columns 2 and 3, guaranteed loans that are uncollateralized are used. Standard errors are clustered at the firm level. ***, **, and * denote 1%, 5%, and 10% statistical significance, respectively (*P*-values are reported in parentheses).

| | Guaranteed loans | Uncollateralized guaranteed loans | |
|---|---|---|---|
| | (1) | (2) | (3) |
| Treatment = 1 | 0.088 | 0.117 | 0.329 |
| | (0.218) | (0.478) | (0.863) |
| Internal credit rating | −0.320*** | −0.342*** | −0.342*** |
| | (0.000) | (0.000) | (0.000) |
| Treatment = 1 × Internal credit rating | | | −0.033 |
| | | | (0.906) |
| Control for firm characteristics | Yes | Yes | Yes |
| Control for loan characteristics | Yes | Yes | Yes |
| Bank–quarter–country–industry FE | Yes | Yes | Yes |
| Observations | 44,422 | 3,793 | 3,793 |
| Adjusted $R^2$ | 0.49 | 0.56 | 0.56 |

notion. Overall, the results in Table VIII show that the selection channel is not a challenging issue in my analysis.

## 5.5 Shock to Asset Values and the Choice of Third-Party Credit Guarantee and Collateral

The collateral channel implies that the value of the assets that a firm has available to pledge as collateral affects the firm's borrowing behavior. Focusing on real estate assets, Luck and Santos (2019) show that following a rise in the value of their real estate assets, firms are more likely to borrow. Further, they argue that smaller firms are more reliant and benefit more from pledging collateral in general and real estate in particular when compared with larger firms. The reason is that smaller firms are more prone to moral hazard problems because they are riskier and more informationally opaque. Hence, smaller firms benefit more from collateral (also consistent with Berger and Udell, 1990; Boot and Thakor, 1994; Rampini and Viswanathan, 2010).

If a third-party credit guarantee is an alternative to collateral, it is expected that borrowers are more likely to rely on third-party credit guarantees and less likely to rely on collateral when they face a negative shock to the value of their assets.

To examine, I take advantage of the information available in FR Y-14Q data, which show the type of collateral used in a loan contract. Specifically, I focus on real estate collateral (constituting less than 20% of collateral types and 13% of all loan observations) for a number of reasons: although the value of the specific collateral pledge to each loan is not known, I can see county-level real estate prices, and their changes over time from public sources of data. Moreover, FR Y-14Q data provide information on the location of borrowing firms; therefore, I can find what county the firm is located. Assuming the value of a borrowing firm's real estate collateral is correlated with the local housing price index, as argued by Luck and Santos (2019), I will be able to investigate whether a negative or positive shock to housing prices affects a firm's choice of real estate collateral versus a third-party credit guarantee.

The source of housing price data is the Federal Housing Finance Agency (FHFA),[31] which includes annual change in county-level housing price indices since 1986 (Bogin, Doerner, and Larson, 2019). I merge county-level housing price data from the FHFA with my loan-level data based on the location of the borrowing firm. An average borrower in the sample has experienced a 3.1% annual growth in the housing prices with the median being 3.3%. The lowest annual growth for a borrower is –26.24% with the highest being 34.95%. For each year, I assume a borrowing firm has experienced a negative (positive) shock to the value of its real estate assets if the firm is in the bottom (top) quartile of annual county price index growth. Further, to investigate the importance of collateral and credit guarantee for smaller firms, I build an indicator that equals 1 if the borrower is in the bottom quartile of the firm size (total assets) across all the borrowers in the sample. As a robustness check, I use the bottom decile as an alternative indicator and I find similar results.

The results are reported in Table IX. In Table IX, Column 1, I examine the relation between the likelihood of using a real estate collateral in a new loan and the growth in the housing price index in the borrowing firm's county, controlling for firm and loan characteristics as well as for the bank–quarter–country–industry fixed effects. The data include loan observations for which county-level price data are available in the FHFA (i.e., only for US borrowers). The results show that firms in the top quartile of real estate price growth—that is, firms that experience a positive shock in the value of their real estate assets—are significantly more likely to pledge a real estate collateral for their new borrowing. The coefficient for the indicator of positive shock is 0.034, significant at the 1% level, indicating that being in the top quartile of real estate price growth increases the likelihood of pledging real estate assets as collateral by 3.4 percentage points (as mentioned above, on average 13% of loans have real estate collateral).

In Column 2, I interact indicators of positive and negative real estate price shocks with the indicator for small firms. Supporting the results in Column 1, the results in Column 2 show that an average firm that experiences a positive shock is 3.1 percentage points more likely to pledge real estate collateral (significant at the 1% level) relative to other borrowing firms. As a point of comparison, about 9.3% of loans have real estate collateral. More importantly this likelihood increases by 1.1 percentage points if the firm that has experienced a positive shock is a small firm. The coefficient is significant at the 10% level. Furthermore, the results show that smaller firms that experience a negative shock to the value of their real estate assets are 1.3 percentage points less likely to pledge real estate collateral in their new loans. The results are robust to the definition of small firm (e.g., using below median size as the measure of small firm). These findings support the notion that collateral plays a critical role in firms' borrowing decisions. Specifically, collateral is more valuable for smaller firms.

31 Available at https://www.fhfa.gov/DataTools/.

**Table IX.** Response to shocks to real estate assets

This table investigates the relationship between the likelihood of using a real estate collateral and a third-party credit guarantee in a new loan following a shock to real estate prices. The unit of observation in each regression is a loan. The sample covers all C&I loans issued by banks during 2012–18. A firm is considered to be affected by a negative (positive) shock to real estate assets if the borrower is located in a county in the bottom (top) quartile of real estate annual price growth across all counties in the year of loan initiation. Standard errors are clustered at the firm level. ***, **, and * denote 1%, 5%, and 10% statistical significance, respectively (*P*-values are reported in parentheses).

| | Real estate collateral | | Credit guarantee |
|---|---|---|---|
| | (1) | (2) | (3) |
| Bottom quartile of real estate price growth = 1 | –0.003 | –0.000 | –0.011 |
| | (0.308) | (0.905) | (0.102) |
| Top quartile of real estate price growth = 1 | 0.034*** | 0.031*** | 0.002 |
| | (0.000) | (0.000) | (0.682) |
| Small firm = 1 | | –0.001 | 0.086*** |
| | | (0.863) | (0.000) |
| Bottom quartile of real estate price growth = 1 × Small firm = 1 | | –0.013** | 0.052*** |
| | | (0.046) | (0.000) |
| Top quartile of real estate price growth = 1 × Small firm = 1 | | 0.011* | –0.001 |
| | | (0.091) | (0.884) |
| Control for firm characteristics | Yes | Yes | Yes |
| Control for loan characteristics | Yes | Yes | Yes |
| Bank–quarter–country–industry FE | Yes | Yes | Yes |
| Observations | 68,633 | 68,633 | 68,633 |
| Adjusted $R^2$ | 0.17 | 0.17 | 0.24 |

The results in Columns 1 and 2 prove novel but expected results considering the vast literature that exists on collateral. In Column 3, I use the same methodology as in Column 2, but this time I investigate the likelihood of using a third-party credit guarantee as the dependent variable. If a third-party credit guarantee substitutes collateral, I would expect to see that small firms are more likely to use a third-party credit guarantee when they experienced a negative shock to the values of their assets. The results in Column 3 are in line with this expectation. The results show that smaller firms in general are 8.6 percentage points more likely to use a third-party credit guarantee. The coefficient for the small firm indicator is significant at the 1% level. Moreover, the likelihood of using a third-party credit guarantee increases by 5.2 percentage points if the firm experiences a negative shock to the value of its asset (significant at the 1% level).

These findings highlight the importance of third-party credit guarantees in firms' borrowing behavior. Although pledging their assets as collateral is instrumental in firms' borrowing decisions (and by extension to their investments), firms also rely on third-party credit guarantees as another important tool to obtain credit. This tool is particularly important for smaller firms, which are the sources of entrepreneurship and employment opportunities and drivers of local economic growth.

## 6. Conclusion

This study addresses five questions. First, how pervasive is the use of third-party credit guarantees by US banks? Second, how do third-party credit guarantees affect a loan's performance and bank estimates of LGD? Third, how do third-party credit guarantees affect the cost of debt for borrowing firms? Fourth, what is the substitutability of third-party guarantees and collateral? Fifth, how does the source of a credit guarantee affect the borrowing firm?

The results of this study are particularly important as they contrast prior emphasis that has been put on collateral. Providing evidence that third-party credit guarantees are prevalent, this study shows that third-party credit guarantees are particularly important for smaller firms often understudied in the empirical literature. Using an identification strategy that controls for bank-, time-, and borrowing firm unobserved risk, I find that the presence of a credit guarantee is associated with a significantly lower interest rate for the guaranteed loan than an otherwise similar unguaranteed loan. Examining whether the type of credit guarantee is related to the price effect, the results show that credit guarantees provided by US government agencies have the largest effects on interest rates. Corporate guarantees offered by the borrowing firm's parent company are smaller but still economically significant to the price. I do not find consistent evidence that guarantees provided by managers or owners (personal guarantees) significantly affect interest rates. Banks are willing to provide a discount on guaranteed loans because credit guarantees provide a risk-mitigating effect. This phenomenon is confirmed further by another finding: Guaranteed loans on average have lower bank estimated LGD ratios and are less likely to become nonperforming when compared with similar loans that are not guaranteed. Other results in this study indicate that a third-party credit guarantee has a significant effect beyond the security that a collateral provides. Third-party credit guarantees and collateral are partially substitutes; however, each also provides specific risk-mitigation benefits that are complementary to each other. Moreover, a negative shock to asset values increases the likelihood that small firms use credit guarantees rather than collaterals to assure banks of loan repayments.

## Data Availability Statement

The data underlying this article were provided by the Federal Reserve by permission. Data will be shared on request to the corresponding author with permission of the Federal Reserve.

## Appendix A: Variable Definitions

This table presents the definitions for the variables used in the paper. The item numbers refer to data fields on Schedule H1 of the FRY-14Q data, available at https://www.federalreserve.gov/reportforms/forms/FR_Y-14Q20181231_i.pdf

| Variable | Definition |
|---|---|
| Panel A: Loan characteristics | |
| Amount (million USD) | Amount of loan outstanding in million dollars (item #25). |
| Credit Guarantee | Indicator variable equal to 1 if the loan is fully or partially guaranteed. The indicator is constructed based on the field Guarantor Flag (item #44), which reflects the type of guarantee on a loan. The allowable values for this field are 1 (fully guaranteed), 2 (partially guaranteed), 3 (fully guaranteed by a U.S. government agency), and 4 (no guarantee). The indicator equals to one if Guarantor Flag equals 1, 2, or 3, and equals to zero if Guarantor Flag equals 4. |
| Credit Guarantee–Agency | Indicator variable equal to 1 if the loan is fully guaranteed by a US government agency. The variable is based on item #44, item #46, and item #4. |
| Credit Guarantee–Corporate | Indicator variable equal to 1 if the loan is guaranteed by the borrower's parent or group member. The variable is based on item #44, item #46, and item #4. |
| Credit Guarantee–Personal | Indicator variable equal to 1 if the loan is guaranteed by an individual (natural person). The variable is based on item #44, item #46, and item #4. |
| Fixed interest rate | Indicator variable equal to 1 if the loan is a fixed-rate loan (item #37). |
| Interest rate (prc) | Effective annual interest rate charged on the loan (item #38). Interest rate is reported in percentage. |
| LGD | LGD as estimated by the lending bank, which gives the percentage of exposure the bank might lose in case the borrowing firm defaults (item #89). |
| Maturity (months) | Number of months between loan origination date (item #18) and maturity date (item #19). |
| Past due | Indicator variable equal to 1 if the number of days principal and/or interest payments past due are equal to or greater than 30 days (item #32). |
| Collateral | Indicator variable equal to 1 if security is provided by collateral (item #36). |
| Syndicated | Indicator variable equal to 1 if the loan is considered syndicated (item #34). |
| Term loan | Indicator variable equal to 1 if the loan belongs to a term loan category (item #20). |
| Panel B: Firm characteristics | |
| Domestic | Indicator variable equal to 1 if the borrowing firm is headquartered in the USA (item #6). |
| Firm size (log(assets)) | Natural logarithm of borrower's book value of total assets (item #70). |
| Internal credit rating | Credit rating that the lending bank has assigned to the borrowing firm (item #10) converted to a 10-grade S&P ratings scale. Each bank is required to disclose its internal risk rating together with a concordance mapping to a 10-grade S&P scale ranging from |

(continued)

. Continued

| Variable | Definition |
|---|---|
| | AAA to D, with the highest possible grade being 10 (equivalent to a AAA rating). |
| Leverage (long-term debt/assets) | Book value of long-term debt (item #78) divided by book value of total assets (item #70). |
| Liquidity (current ratio) | Book value of current assets (item #66) divided by book value of current liabilities (item #76) |
| Profitability (net income/sales) | Net income (item #59) divided by sales (item #54). |
| Tangibility (tangible assets/assets) | Book value of tangible assets (item #68) divided by book value of total assets (item #70) |

## Appendix B: Summary Statistics for Main Tests

The table summarizes the loan-level data from Federal Reserve Y-14Q reporting forms. The sample period is from 2012 to 2018. The sample covers USD commercial and industrial loans with valid information originated during the sample period by large US bank holding companies. The sample is further restricted to loans that are used in regression setups, where borrowing firm and lender time-varying fixed effects are controlled. In order to prepare the sample, I start with the original sample of 148,590 loans, as presented in Table I, and restrict the sample to only loans for which there is at least one other loan in the sample, where the borrowing firm, the lending bank, and the quarter of originations are the same (24,573 loans). All variables are defined in Appendix A.

Summary statistics at loan level

| | Mean | SD | 10% | Median | 90% |
|---|---|---|---|---|---|
| Credit guarantee | 0.424 | 0.494 | 0.000 | 0.000 | 1.000 |
| Fully guaranteed | 0.378 | 0.485 | 0.000 | 0.000 | 1.000 |
| Interest rate (prc) | 3.291 | 1.584 | 1.650 | 3.117 | 5.220 |
| Amount (million USD) | 13.643 | 50.269 | 1.000 | 4.511 | 30.300 |
| Log(amount) | 15.339 | 1.496 | 13.816 | 15.322 | 17.227 |
| Maturity (months) | 52.741 | 29.241 | 12.000 | 60.000 | 82.000 |
| Log(maturity) | 3.779 | 0.711 | 2.485 | 4.094 | 4.407 |
| Collateral | 0.773 | 0.419 | 0.000 | 1.000 | 1.000 |
| Fixed interest rate | 0.264 | 0.441 | 0.000 | 0.000 | 1.000 |
| Term loan | 0.474 | 0.499 | 0.000 | 0.000 | 1.000 |
| Syndicated loan | 0.440 | 0.496 | 0.000 | 0.000 | 1.000 |
| LGD (prc) | 36.563 | 12.923 | 20.000 | 38.000 | 50.000 |
| Past due (prc) | 2.592 | 15.889 | 0.000 | 0.000 | 0.000 |
| Domestic | 0.905 | 0.293 | 1.000 | 1.000 | 1.000 |
| Internal credit rating | 6.072 | 0.983 | 5.000 | 6.000 | 7.000 |
| Firm size (log(assets)) | 19.135 | 2.474 | 16.194 | 18.895 | 22.514 |
| Leverage (long-term debt/assets) | 0.324 | 0.250 | 0.028 | 0.285 | 0.635 |
| Profitability (net income/sales) | 0.051 | 0.121 | –0.011 | 0.038 | 0.145 |
| Liquidity (current ratio) | 1.675 | 1.611 | 0.559 | 1.302 | 2.879 |
| Tangibility (tangible assets/assets) | 0.795 | 0.246 | 0.400 | 0.912 | 1.000 |